\documentclass[english]{article}

\usepackage{amsthm}
\usepackage{amsmath}
\usepackage{graphicx}
\usepackage{amssymb}
\usepackage{fullpage}

\begin{document}
\title{Non-Riemannian metric emergent from scalar quantum field theory}
\author{Arnab Kar\footnote{arnabkar@pas.rochester.edu}, S. G. Rajeev\footnote{rajeev@pas.rochester.edu} \footnote{Also at the Department of Mathematics}}
\maketitle
\begin{center}
Department of Physics and Astronomy\\ 
University of Rochester\\
Rochester, New York 14627, USA
\end{center}

\begin{abstract}
We show that the two-point function $\sigma(x,x')=\sqrt{\langle[\phi(x)-\phi(x')]^{2}\rangle}$
of a scalar quantum field theory is a metric (i.e., a symmetric positive
function satisfying the triangle inequality) on space-time (with imaginary
time). It is very different from the Euclidean metric $|x-x'|$ at large distances, 
yet agrees with it at short distances. For example, space-time has finite diameter which is not universal.
The Lipschitz equivalence class of the metric is independent of the cutoff.
 $\sigma(x,x')$ is not the length of the geodesic in
any Riemannian metric. Nevertheless, it is
possible to embed space-time in a higher  dimensional space  so that $\sigma(x,x')$ is the length of the geodesic in
the ambient space. $\sigma(x,x')$ should be useful in constructing the continuum limit of
quantum field theory with fundamental scalar particles.

\begin{description}
\item[PACS numbers] 11.10.Cd,\, 11.10.Gh,\, 14.80.Bn,\, 84.37.+q
\end{description}

\end{abstract}

\maketitle

\section{Introduction}
In view of the anticipated discovery of the Higgs et al. boson\cite{Higgsetal}, 
it is timely to reconsider the fundamental  implications of a plain scalar field of 
the standard model: one  that is not composite or associated to  supersymmetry 
or  to extra dimensions of space-time. It is of interest to study a quantum theory 
of scalar fields in any case, as it describes many other phenomena, such as phase transitions.

Such a field is not expected to have an effect on the geometry of space-time. 
This is unlike the other bosonic fields: according to general relativity, gravity 
modifies the metric of space-time from Euclidean to Riemannian. And gauge 
fields have a geometric meaning in terms of parallel transport. A fundamental 
scalar field is thought to have a geometrical meaning only as a remnant of dimensional 
reduction: if space-time has extra dimensions (whether continuous or discrete, 
as in noncommutative geometry), the extra components of the gauge or gravitational field 
would be scalars in four-dimensional space-time.

In this paper, we will show that a scalar quantum field defines a 
metric on space-time as well. But to understand this metric we must 
go beyond Riemannian geometry. In recent years, the study of such 
general metric spaces has emerged\cite{Burago}  
as a fundamental branch of mathematics, touching on topology, geometry and analysis.

In classical mechanics, a free particle moves along a straight line. The length of 
this line is the distance between points. In quantum mechanics, the propagator 
is the sum over all paths\footnote{In this paper we will study quantum theories \cite{GlimmJaffe} formulated
in terms of a path integral $\int e^{-S}\mathcal{D}\phi$ where $S$
is a real positive function of the field, the action. That is why
the signature of the metric is positive instead of being Lorentzian.
To get physical answers we must do an analytical continuation in time.},
with a weight proportional to $e^{-S}$, where $S={1\over 2}\int \dot{x}^2dt$ is the action. 
Thus, a quantum notion of distance should involve the propagator itself, rather than a property of a particular path.

In quantum field theory, we should seek a notion of distance based on 
the correlations of fields at two points, which is the analogue of the propagator. 
We will show that the quantity
\[
\sigma(x,x')=\sqrt{\langle [\phi(x)-\phi(x')]^{2}\rangle}
\]
(defined with a regularization) in a scalar quantum field theory satisfies the 
axioms of a metric \cite{Burago} (see the Appendix for a summary of metric 
geometry). In particular, the triangle inequality
\[
\sigma(x,x')\leq \sigma(x,x'')+\sigma(x'', x')
\]
holds. This notion of distance between two points is very different, 
from the Riemannian  notion. It is worth exploring on its own right, even if we continue to use the Euclidean length for most physical purposes.

For example, the triangle inequality above cannot  be saturated if $x,x',x''$ 
are all distinct. By contrast, for the Euclidean distance, as long as $x''$ lies 
on the straight line connecting $x$ to $x'$,  we would have  $|x-x'|=|x-x''|+|x''-x'|.$ 
More generally, in Riemannian geometry, we can saturate the triangle 
inequality by choosing $x''$ to be any point on the shortest geodesic connecting $x$ to $x'$.

This means that $\sigma(x,x')$ is not the length of geodesics in any 
Riemannian geometry: it defines a non-Riemannian metric geometry.

So why would space-time look Euclidean classically? It turns out that 
the length of a curve defined by $\sigma(x,x')$ (in a superrenormalizable 
or asymptotically free theory; we do not know the general answer yet) is the 
same as the Euclidean length. The reason is that for small distances and 
small interactions, $\sigma(x,x')\propto |x-x'|$. (The  proportionality constant 
the depends on the regularization.) The length of a curve is defined by 
breaking it up into small segments (see the Appendix). For small enough 
segments we will get (up to a constant) their  Euclidean length. Classical 
measurements of distance always involve lengths of curves. So 
even $\sigma$ will give the Euclidean answer (up to  the proportionality constant) in these measurements.

According to $\sigma$,  the shortest curve connecting two points is still 
the straight line. But this shortest length  is not the same as $\sigma(x,x')$. 
We will see that $\sigma(x,x')$ is a monotonically   increasing function of 
$|x-x'|$, which for  dimensions $n=3,4$ tend to a constant for large $|x-x'|$: 
space-time has finite diameter according to $\sigma$.
For the case of a massless free field in four dimensions using the heat kernel 
regularization, we have an explicit formula:
\[
\sigma(x,x')=\left[\frac{1}{8 \pi ^2 a^2}-\frac{1}{2 \pi ^2 (x-x')^2} +\frac{e^{-\frac{(x-x')^2}{4 a^2}}}{2 \pi ^2 (x-x')^2}\right]^{1\over 2}
\]

The story can be different for  a scalar quantum field theory that does 
not tend to free field theory at short distances. For a $\lambda\phi^4$ 
theory in four dimensions, perturbation theory breaks down at short 
distances. We are therefore, not able to determine analytically the 
relationship of $\sigma(x,x')$ to $|x-x'|$ for short distances. This case is 
of great importance, as it describes the Higgs boson of the standard model. 
Large-scale computer simulations are needed to study this relationship. 
Even if the Higgs boson turns out to be a fundamental particle and there 
are no indications of supersymmetry, compositeness or extra dimensions, 
the LHC will serve as an exciting as a probe of  this non-Riemannian metric of space-time.

If the metric depends on the cutoff $a$, can it still have a physical significance? 
We will see, in the continuum regularization schemes, a change of the cutoff 
does not change the Lipschitz equivalence class of the metric. Thus, the 
Lipschitz class of space-time should have a physical significance: instead 
of differentiable functions, we would talk of Lipschitz functions, for example. 
This equivalence class does change if we let the cutoff go to zero: it is different 
from that of Euclidean space. Thus, the functions that are Lipschitz with 
respect to $\sigma$ are not the same as those with respect to $|x-x'|$.

If we use discrete  regularization schemes (e.g., lattice), the correct notion of 
equivalence of metrics might be quasi-isometry. Gromov\cite{Gromov} 
used such a notion to show that groups of polynomial growth are discrete 
approximations to Euclidean space.

When non-Euclidean geometry was still new, it was useful to understand
a curved metric in terms of an embedding into Euclidean space. In
the same way, it is useful to understand a non-Riemannian metric such
as ours by embedding into a Riemannian manifold.  We will show that our 
metric $\sigma(x,x')$ can be thought of as the length of the geodesic in a 
Riemannian manifold with one extra dimension: the geodesic does not lie 
in the submanifold, so it has a shorter length than any curve that stays within 
the submanifold (in particular the Euclidean straight line).

In section 2, we describe scalar quantum field theory on a lattice and 
in section 3, how $\sigma(x,x')$ can be defined for it. We then calculate 
the metric explicitly for the case of a free massless scalar field on a lattice in terms 
of certain discrete Fourier series. This special case is related to the resistance 
metric of Kigami. We show in section 4 that the metric cannot be 
induced by any Riemannian geometry. In section 5 we consider other 
regularization schemes, in particular the heat kernel method. This allows us 
to get an explicit form in terms of elementary functions for  free field theory. 
It is shown that the Lipschitz equivalence class of the metric is independent 
of the cutoff. A first step toward understanding interacting theories is taken in section 6 
where we calculate the metric in the $\epsilon$-expansion of Wilson and Fischer 
fixed point of critical phenomena. In section 7, we show that it is possible to 
embed space-time in a Riemannian manifold of one higher dimension such 
that our $\sigma$ is the induced metric. In the last section, we summarize our 
results and give some directions for further research. And finally, in an Appendix, 
we collect together some facts about metric geometry that are known 
in the mathematics literature but rarely used by physicists.

\section{Lattice Scalar Field Theory}

A scalar field on a graph $\Gamma$ is a function $\phi:\Gamma\to \mathbb{R}^{N}$.
The action (or energy, depending on the physical application) is
a function of the scalar field

\begin{equation}
S(\phi)=\frac{a^{n-2}}{2}\sum_{x\sim x'}\left[\phi(x)-\phi(x')\right]^{2}+a^{n}\sum_{x}V(\phi(x)) \label{eq:Sphi}
\end{equation}

The first sum is over nearest neighbors in the graph and $a$ is the
distance between them\footnote{Obviously, we can absorb $a$ into $\phi$ or $V$, but we will find
it convenient not to do so.}.
$V$ is a polynomial whose coefficients are the ``bare coupling
constants''. Free field theory is the special case where $V$ is a quadratic function. Massless free field theory is the case $V=0$. The
expectation value of any function of the field is defined to be
\begin{equation}
\langle f \rangle=\frac{\int f(\phi)e^{-S(\phi)} d\phi}{\int e^{-S(\phi)} d\phi}
\end{equation}

In particular, the correlation functions are the expectation values
\[
G(x_{1},\cdots x_{n})=\langle \phi(x_{1})\cdots\phi(x_{n})\rangle \label{eq:CorrelationFns}
\]

Sometimes it will be more convenient to work with quantities such as
\[
R(x,x')=\langle[\phi(x)-\phi(x')]^{2}\rangle
\]

related to the correlation function.

\begin{equation}
R(x,x')=G(x,x)+G(x',x')-2G(x,x').\label{eq:RvsG}
\end{equation}

The case of greatest interest\cite{Creutz} is a cubic lattice $\Omega_{a,L}=a\left(\mathbb{Z}/\Lambda\mathbb{Z}\right)^{n}$
with period $L=\Lambda a$. The aim of quantum field
theory is to construct the continuum limit $a\to0,L\to\infty$ such
that the correlation functions have a sensible limit. In taking this
limit, the coefficients of the polynomial $V$ are to be varied as
functions of the cutoff $a$. This program is essentially complete \cite{GlimmJaffe} 
in the case $n=2$. Constructing nontrivial examples
(i.e., with a $V$ of degree higher than two) is very difficult in
the physically interesting cases of dimensions three (for the theory
of phase transitions) and four (for particle physics). Wilson and
Fischer used ingenious approximations\cite{Fisher, ZinnJustin} to understand
the three-dimensional case. In dimensions higher than four, such a
continuum limit does not exist except for the case of a free field\cite{NoGoTheorem}. 
The case of four dimensions is marginal, and a
nontrivial continuum limit cannot be constructed by standard methods\cite{FernandezFroehlichSokal}.

\section{Standard Deviation Metric}

The mean deviation
\[
D(x,x')=\langle|\phi(x)-\phi(x')|\rangle
\]
satisfies the triangle inequality since, for each instance of $\phi$,
\[
|\phi(x)-\phi(x')|\leq|\phi(x)-\phi(x'')|+|\phi(x'')-\phi(x')|
\]
holds. So it will hold in the average as well.

More generally
\[
D_{p}(x,x')=\left[\langle|\phi(x)-\phi(x')|^{p}\rangle\right]^{\frac{1}{p}}
\]
for $p\geq1$ is a metric on space-time\footnote{Due to
translation invariance, $\langle |\phi(x)-\phi (x')|\rangle=0$.}. The most interesting is the
case $p=2$ of the standard deviation
\begin{equation}
\sigma(x,x')=\left[\langle|\phi(x)-\phi(x')|^{2}\rangle\right]^{\frac{1}{2}}.
\end{equation}

$\sigma(x,x')$ is obviously positive and symmetric. Also, $\sigma(x,x')>0$
if $x\neq x'$ because, otherwise, every instance of a scalar field
would have to take the same value at $x$ and $x'$.

Each theory of matter field will define a metric on space-time.
The distance is a simple concept for scalar fields. For gauge fields,
it is more subtle, but gauge invariant notions do exist \cite{FeynmanYM3}.
Reflection positivity seems to imply such a metric even for fermion
fields. When the scalar field
takes values in a curved target space $\phi:\Omega^{n}\to M$ (e.g.,
the nonlinear sigma model \cite{ZinnJustin}), we would use the metric
$d_{M}$ of the target to define $\sigma(x,x')=\sqrt{\langle|\left[d_{M}(\phi(x),\phi(x'))\right]^{2}|\rangle}$.

Why could we not have defined a metric using the average of the square
of the distance
\[
R(x,x')=\langle(\phi(x)-\phi(x'))^{2}\rangle
\]
itself? It is more closely related to the correlation functions (\ref{eq:RvsG}).
The point is that the square of a metric, such as $(\phi(x)-\phi(x'))^{2}$,
does not in general satisfy the triangle inequality. (By contrast,
the square root of a metric always does.) In some special cases
(e.g., massless free field) the expectation value $R(x,x')$, itself,
is a metric. This particular case was discovered by Kigami in the
context of fractals \cite{Kigami,Strichartz}. But there are other
probability distributions (that are not Gaussians) for which $R(x,x')$
does not satisfy the triangle inequality. Also, with regularizations
other than the lattice (e.g., heat kernel method, see section below),
$R(x,x')$ does not satisfy the triangle inequality. But $D_{1}(x,x')$
and $\sigma(x,x')$ always do. Of the two, the standard deviation 
 $\sigma(x,x')$ is easier to calculate, as usual.

\section{Free Scalar Field}

In the special case(\ref{eq:Sphi}), of a free field
\begin{equation}
V(\phi)=\frac{1}{2}m^{2}\sum_{x}\phi(x)^{2},
\end{equation}

$\phi(x)-\phi(x')$ is a Gaussian random variable with variance $R(x,x')$
and zero mean. So we have the relation

\begin{equation}
\sigma=\sqrt{\frac{\pi}{2}}D.
\end{equation}

This relation is universal for free fields: it does not depend on
the cutoff procedure used (e.g., square vs triangular lattice). But
the range of values of $\sigma$ will depend on the cutoff.\footnote{Note that $I_{k}=\int_{-\infty}^{\infty}|\phi|^{k}e^{-\frac{1}{2}\frac{\phi^{2}}{R}}d\phi=
2R^{\frac{1+k}{2}}\int_{0}^{\infty}e^{-u}[2u]^{\frac{k-1}{2}}du=2^{\frac{k+1}{2}}\Gamma\left(\frac{k+1}{2}\right)R^{\frac{1+k}{2}}$.
Thus $\langle|\phi|\rangle=\frac{I_{1}}{I_{0}}=\frac{2}{\sqrt{2\pi}}\sqrt{R}=
\sqrt{\frac{2}{\pi}R},\langle\phi^{2}\rangle=\frac{I_{2}}{I_{0}}=\frac{2^{\frac{3}{2}}\frac{1}{2}\sqrt{\pi}}{\sqrt{2\pi}}R=R$}

\subsection{Explicit formula for $R$}

The expectation values are Gaussian integrals which we can evaluate
explicitly in terms of the Green's function
\[
G(x,x')=\langle\phi(x)\phi(x')\rangle
\]

It is the solution of the lattice Helmholtz equation,
\[
[\Delta_{x}+m^{2}]G(x,x')=\delta_{\Omega_{a,L}}(x,x').
\]

The lattice Laplacian is the sum over nearest neighbors $y$ for fixed
$x$ of the difference:
\[
\Delta\psi(x)=\frac{1}{a^{2}}\sum_{y\sim x}\left[\psi(x)-\psi(y)\right]
\]
(According to this definition, the eigenvalues of the operator are positive.)

The lattice delta function depends on $L$ through periodicity;
\[
\delta_{\Omega_{a,L}}(x,x')=\begin{cases}
a^{-n},\quad & \mathrm{if}\ x=x'\mathrm{\ mod\ }L\\
0 & \mathrm{otherwise.}\end{cases}
\]

The discrete Fourier transform of a function is given by
\[
\tilde{\psi}(p)=a^{n}\sum_{x\in\Omega_{a,L}}e^{-ip\cdot x}\psi(x),
\]

where the wave number $p$ belongs to the dual lattice
\[
p\in\tilde{\Omega}_{a,L}=\Omega_{\frac{2\pi}{L},\frac{2\pi}{a}}
\]

for which $a,L$ are exchanged for their reciprocals. Note the identity
\[
L^{-n}\sum_{p\in\tilde{\Omega}_{a,L}}e^{ip\cdot(x-x')}=\delta_{\Omega_{a,L}}(x,x').
\]

The inverse discrete Fourier transform is 
\[
\psi(x)=L^{-n}\sum_{p}e^{ip\cdot x}\tilde{\psi}(p)
\]

then
\[
\widetilde{\Delta\psi}(p)=\tilde{\Delta}(p)\tilde{\psi}(p),\quad\tilde{\Delta}(p)=\frac{4}{a^{2}}\sum_{r=1}^{n}\sin^{2}\frac{a p_{r}}{2}.
\]

thus
\[
G(x,x')=L^{-n}\sum_{p\in\tilde{\Omega}_{a,L}}\frac{e^{ip.(x-x')}}{\tilde{\Delta}(p)+m^{2}}
\]

It follows that
\begin{equation}
R(x,x')=L^{-n}\sum_{p\in\tilde{\Omega}_{a,L}}\frac{4\sin^{2}\frac{p\cdot(x-x')}{2}}{\tilde{\Delta}(p)+m^{2}}
\end{equation}

\subsection{The resistance metric}

In the limit $m\to0$ of a free massless scalar field, $R(x,x')$,
itself, (and not only its square root) is a metric. Explicitly
\[
R(x,x')=L^{-n}\sum'_{p\in\tilde{\Omega}_{a,L}}\frac{4\sin^{2}\frac{p.(x-x')}{2}}{\tilde{\Delta}(p)}
\]
the sum being over terms with $\tilde{\Delta}(p)\neq0$. This is the resistance metric
of Kigami \cite{Kigami}, evaluated for the cubic lattice. There is
a simple physical interpretation for this quantity. Imagine a network,
each pair of nearest neighbors being connected by a resistor of same
resistance. Then $R(x,x')$ is the effective resistance between the
pair of points $x,x'$ after all the others have been eliminated using
Kirchoff's laws of current conduction. It is obvious that $R(x,x')$
is positive and symmetric. See Ref. \cite{Strichartz} for an ingenious
proof that it satisfies the triangle inequality. However, we will
not use this fact.

\subsection{The infinite lattice}

The resistance metric is often studied\cite{Strichartz} on sequences
of graphs that tend to a fractal. It has not been studied as a metric
on the more familiar graph of a cubic lattice. Perhaps the reason
is that it is totally different from the Euclidean metric. Physicists\cite{LatticeResistance}
have already calculated the properties of resistance on cubic lattices
without noting that it satisfies the triangle inequality. In the limit
of a lattice of infinite period $L\to\infty$, the momentum space
becomes a torus of period $\frac{2\pi}{a}$. Then, we have the integral
representation\footnote{Use the Euler-MacLauren formula 
$\displaystyle{\lim_{L\to\infty}}L^{-n}\sum_{k=1}^{\Lambda}f\left(\frac{2\pi}{L}k\right)=\int_{0}^{\frac{2\pi}{a}}f(p)\frac{d^{n}p}{(2\pi)^{n}}$
} with $m=0$,

\begin{eqnarray}
R(x,x')&=&\left(\frac{1}{2\pi}\right)^{n}\int_{-\frac{\pi}{a}}^{\frac{\pi}{a}}d^{n}p\frac{4\sin^{2}\frac{p\cdot(x-x')}{2}}{\frac{4}{a^{2}}{\displaystyle {\sum_{r=1}^{n}}\sin^{2}\frac{p_{r}a}{2}}} \\ \nonumber
R(x,x')&=&\frac{a^{2-n}}{n},\quad|x-x'|=a,\quad m=0. \\ \nonumber
\end{eqnarray}

If we consider $x-x'=(0,\cdots a,\cdots0)$ to be along the $i$ th
direction alone and consider the sum over all the $n$ integrals,
the trigonometric terms in the numerator and denominator would cancel
and the integral evaluates to $a^{2-n}$. Also, the nearest neighbors
have the same resistance.

In the opposite limit of large Euclidean distance, the answer depends
more dramatically on the dimension.

For $n=1$, it is easy to see that the resistance metric is simply
the Euclidean distance
\[
R(x,x')=|x-x'|
\]

We have
\[
R(x,x')=\frac{1}{2\pi}\int_{-\frac{\pi}{a}}^{\frac{\pi}{a}}dp\frac{\sin^{2}\frac{p(x-x')}{2}}{a^{-2}\sin^{2}\frac{p a}{2}}
\]

Let $p\to\frac{2\pi}{a}p$ and $r=\frac{|x-x'|}{a}$
\[
R(x,x')=a\int_{-\frac{1}{2}}^{\frac{1}{2}}\frac{\sin^{2}\pi r}{\sin^{2}\pi p}dp=ra=|x-x'|.
\]

For $n=2$,
\[
R(x,x')\to\frac{1}{2\pi}\left[\log\frac{|x-x'|}{a}+\gamma+\frac{1}{2}\log8+\cdots\right],\quad|x-x'|\to\infty
\]

For $n>2$, and $|x-x'|>>a$ we can approximate
\[
R(x,x')\approx\frac{C_{n}}{a^{n-2}}-2G_{n}(x-x')
\]

where $G_{n}(x)$ is the continuum Green's function of the Laplace
operator. The constant $C_{n}$ is independent of $a$ but depends
on the method of regularization. For the lattice regularization,
\[
C_{n}=2\int_{-\frac{1}{2}}^{\frac{1}{2}}\frac{1}{4\left[\sin^{2}\pi p_{1}+\sin^{2}\pi p_{2}+\cdots\sin^{2}\pi p_{n}\right]}d^{n}p
\]

\[
R(x,x')\approx\frac{C_{n}}{a^{n-2}}-\frac{C_{n}'}{|x-x'|^{n-2}}+\mathrm{O}\left(|x-x'|^{n-3}\right)
\]

The constant $C_{n}'$ is universal: it is the same in every regularization
scheme, being simply related to the continuum Green's function.
\[
C_n'={1\over 2}\pi^{-{n\over 2}}\Gamma\left( {n\over 2}-1  \right)
\]

When $n=3$,
\[
C_{3}\approx0.505462,\quad C_{3}' = \frac{1}{2\pi}
\]

For $n=4$ also
\[
C_{4}=0.309867,\quad C_{4}'=\frac{1}{2\pi^{2}}
\]

The plot shows that $\sigma(x,x')=\sqrt{R(x,x')}$ always grows with the Euclidean distance.

But the rate of growth is very slow for large distances. For nearest
neighbors
\[
\sigma(x,x')=a^{1-\frac{n}{2}}\sqrt{\frac{1}{n}},\quad|x-x'|=a.
\]

And for large distances
\[
\sigma(x,x')\approx\sigma_{n}a^{1-\frac{n}{2}}-\frac{\sigma_{n}'}{|x-x'|^{n-2}}+\mathrm{O}\left(|x-x'|^{n-3}\right)
\]
\[
\sigma_{n}=\sqrt{C_{n}},\quad\sigma_{n}'=a^{\frac{n}{2}-1}\frac{C_{n}'}{2\sqrt{C_{n}}}.
\]

\begin{figure}
\begin{center}
\includegraphics[scale=0.6]{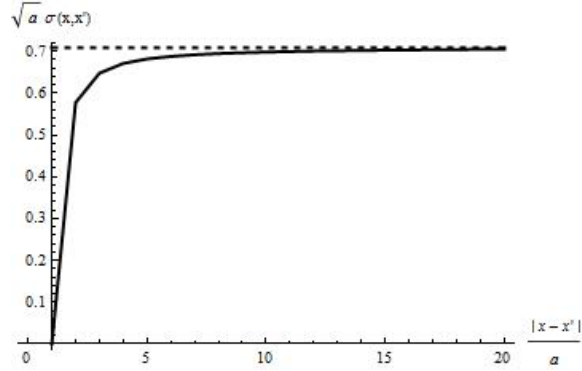}
\end{center}
\caption{\label{sigma3} $\sqrt{a}\sigma(x,x')$ is plotted as a function of Euclidean distance scaled by lattice length.}
\end{figure}

Note that the above equation is consistent with $C_{n}>\frac{1}{n}$.
We plot (Fig.~\ref{sigma3})  the distance in dimension three, in units of the lattice spacing
$a$. Note that the length of a path is the sum of distances between
nearest neighbors along the path. Hence the geodesic distance (the
length of the shortest path connecting two points) is proportional
to the Euclidean distance
\[
\sigma_{l}(x,x')=a^{-\frac{n}{2}}\sqrt{\frac{1}{n}}|x-x'|\geq\sigma(x,x')
\]

The equality holds only for nearest neighbors. For large $|x-x'|$,
the geodesic distance is much greater: $\sigma_{l}(x,x')>>\sigma(x,x')$.
It is easy to understand why using the resistance model, the shortest
curve is just one among possibly many paths that connect the pair
of points. When resistances corresponding to the paths are combined
in parallel, the effective resistance obtained is smaller than all
of them.

\subsection{No intermediate point}

We now show that when $n\geq3$ , the triangle inequality
\[
\sigma(x,x')\leq\sigma(x,x'')+\sigma(x'',x')
\]
can never be saturated unless one of the distances is zero.
The distance between nearest neighbors is
\[
\frac{1}{\sqrt{n}}a^{1-\frac{n}{2}}
\]

So the smallest value for the rhs, being the sum of two nonzero
distances, is twice this:
\[
\frac{2}{\sqrt{n}}a^{1-\frac{n}{2}}.
\]

On the other hand, the distance between any pair of points is bounded:
\[
\sigma(x,x')<\sqrt{C_{n}}a^{1-\frac{n}{2}}
\]

Now, its easy to check that
\[
C_{n}<\frac{4}{n}.
\]

Instead of an analytic proof, we can simply calculate the values in
the two interesting cases numerically:

\begin{eqnarray*}
C_{3}&\approx&0.505462<\frac{4}{3} \\ \nonumber
C_{4}&\approx&0.309867<1. \\ \nonumber
\end{eqnarray*}

Thus, the minimum value of the rhs is greater than the maximum
value of the lhs and the inequality can never be saturated.

It follows that $\sigma(x,x')$ cannot be induced by any Riemannian
metric.

\subsection{Massive scalar field}

In the case of a massive free scalar field, the asymptotic dependence
on the Euclidean distance is given by the Yukawa potential between
points charges exchanging a massive particle: it vanishes exponentially,
\begin{equation}
\sigma(x,x')\approx\sigma_{n}a^{1-\frac{n}{2}}-\sigma_{n}'\frac{e^{-m|x-x'|}}{|x-x'|^{n-2}}+\mathrm{O}\left(|x-x'|^{n-3}\right),\quad n>2.
\end{equation}

The constants $\sigma_{n},\sigma_{n}'$ are as above.

\section{Other Regularization Schemes}

Although we have used the lattice definition of scalar field theories,
universality implies that other regularization schemes will suffice.
For example, we could use a sharp momentum cutoff or a smooth momentum
cutoff or a heat kernel method.

We saw that in the limit $L\to\infty$, the momentum variables take
values on a torus of period $\frac{2\pi}{a}$ in each direction. Thus, the
lattice regularization amounts to replacing a potentially divergent
integral by
\[
\int d^{n}pf(p)\to\int_{-\frac{\pi}{a}}^{\frac{\pi}{a}}\frac{d^{n}p}{(2\pi)^{n}}f'(p)
\]
where $f'(p)$ agrees with $f(p)$ for $|p|<<\frac{1}{a}$. A typical example we encountered above is
\[
\frac{1}{p^{2}}\to\frac{1}{a^{-2} \displaystyle{\sum_{r=1}^{n}}4\sin^{2}\frac{a p_{r}}{2}}
\]

Another method, commonly used in scalar quantum field theory is
\[
\int d^{n}pf(p)\to\int d^{n}pK\left(a|p|\right)f(p)
\]
where $K(\xi)\approx1$ for $\xi<<1$ and is zero for $\xi>>1$. The
advantage is that this preserves rotation invariance which the lattice
breaks. Examples are

\begin{eqnarray*}
K(a|p|)&=&\frac{1}{1+a^{2}p^{2}},\quad \mathrm{Pauli-Villars} \\
K(a|p|)&=&\Theta(a|p|<1),\quad \mathrm{Sharp\ Cutoff} \\
K(a|p|)&=&e^{-a^{2}p^{2}},\quad \mathrm{Heat\ Kernel} \\
\end{eqnarray*}

Polchinski\cite{Polchinski}, among others, has advocated for a smooth
function that is one for $|p|< a^{-1}$ and zero for $|p|>a^{-1}$.

The advantage of these schemes is that the underlying space-time continues
to be $\mathbb{R}^{n}$, but with a possibly different measure of
integration on its dual space (momentum space). Our proposal would
be to determine a metric on space-time from the standard  deviation
computed using this regularized measure in momentum space. Again we
begin with the free field,

\begin{eqnarray*}
G(x,x')&=&\int {d^{n}p\over (2\pi)^n}K(a|p|)\frac{1}{p^{2}+m^{2}}e^{ip\cdot(x-x')} \\
R(x,x')&=&\int {d^{n}p\over (2\pi)^n } K(a|p|)\frac{4\sin^{2}\frac{p.(x-x')}{2}}{p^{2}+m^{2}} \\
\end{eqnarray*}

The explicit answer seems simplest for the heat kernel regularization.
With $m=0$, we get an answer in terms of the incomplete Gamma function,
$\Gamma(\nu,z)=\int_{z}^{\infty}t^{\nu-1}e^{-t}dt$:
\begin{equation}
G(x,x')=\frac{\pi^{\frac{n}{2}}}{4 |x-x'|^{n-2}}\left[\Gamma\left(\frac{n-2}{2}\right)-\Gamma\left(\frac{n-2}{2},\frac{(x-x')^{2}}{4a^{2}}\right)\right]
\end{equation}

To see this,
\begin{eqnarray*}
G(x,x')&=& \int {d^np\over (2\pi)^n}e^{-a^2 p^2}\frac{1}{p^{2}}e^{ip\cdot(x-x')}\\
&=& \int_{a^2}^\infty dt  \int {d^np\over (2\pi)^n}e^{-t p^2} e^{ip\cdot(x-x')}\\
&=&\int_{a^2}^\infty dt { e^{-{(x-x')^2\over 4t}}   \over (4\pi t )^{n\over 2}} \\
\end{eqnarray*}
which we evaluate. $a$ plays the same role as the nearest-neighbor distance in the lattice regularization. It is the short
distance cutoff. 

In particular,
\begin{equation}
G(x,x)=\frac{2^{1-n} \pi ^{-\frac{n}{2}} a^{2-n}}{n-2}
\end{equation}

\begin{equation}
R(x,x')=\frac{2^{2-n}\pi^{-\frac{n}{2}}}{n-2}a^{2-n}-{1\over 2}\frac{|x-x'|^{2-n}}{\pi^{\frac{n}{2}}}\left[\Gamma\left(\frac{n-2}{2}\right)-\Gamma\left(\frac{n-2}{2},\frac{(x-x')^{2}}{4a^{2}}\right)\right]
\end{equation}

For even $n$ the expression is more elementary:
\begin{equation}
R(x,x')=\frac{1}{8 \pi ^2 a^2}-\frac{1}{2 \pi ^2 (x-x')^2} +\frac{e^{-\frac{(x-x')^2}{4 a^2}}}{2 \pi ^2 (x-x')^2},\quad n=4.
\end{equation}

It follows that our metric $\sigma(x,x')=\sqrt{R(x,x')}$ is proportional
to the Euclidean distance for small $|x-x'|$:

\begin{eqnarray*}
R(x,x')&=&\frac{2^{-n}\pi^{-\frac{n}{2}}(x-x')^{2}}{n a^{n}}+O\left(|x-x'|^{4}\right)\\
\sigma(x,x')&=&   \frac{2^{-{n\over 2}}\pi^{-\frac{n}{4}}}{\sqrt{n} a^{n\over 2}}|x-x'|+O\left(|x-x'|^{2}\right)  \\
\end{eqnarray*}

\begin{figure}
\begin{center}
\includegraphics[scale=0.6]{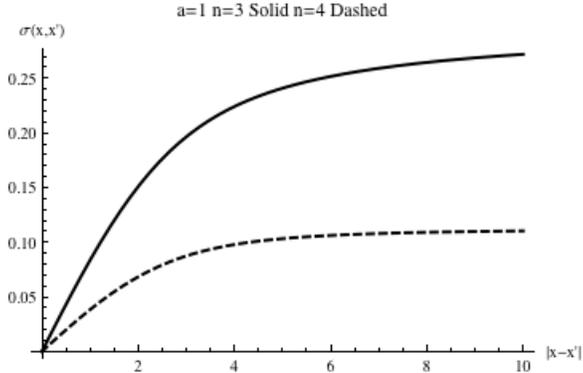}
\end{center}
\caption{\label{sigmaplot}This plot shows the dependence of $\sigma$ with Euclidean distance in dimensions three and four.}
\end{figure}

We plot (Fig.~\ref{sigmaplot}) the metric as a function of Euclidean distance, in units with
$a=1$ for $n=3,4$.

Note that $R(x,x')$ would not satisfy the triangle inequality, being
proportional to the \emph{square} of the Euclidean distance for small
$|x-x'|$. This confirms that the correct choice of metric is the
standard deviation, not the variance of the scalar field.

For large $|x-x'|$,
\begin{equation}
R(x,x')={2^{2-n}\pi^{-{n\over 2}}\over (n-2)a^{n-2}}-\frac{\pi^{-{n\over 2}}}{2}\Gamma\left( {n\over 2}-1  \right){1\over |x-x'|^{n-2}}+\cdots
\end{equation}

Thus, in the heat kernel regularization
\[
C_{n}={2^{2-n}\pi^{-{n\over 2}}\over n-2 } ,\quad C_{n}'={1\over 2}\pi^{-{n\over 2}}\Gamma\left( {n\over 2}-1  \right)
\]

As noted earlier $C_{n}$ is not universal but $C_{n}'$ is. To compare the numerical values
\begin{eqnarray*}
C_{3} &\approx&   0.0897936  \quad\mathrm{vs} \ 0.505462\ \mathrm{for\ lattice} \\
C_{4} &\approx&  0.0126651 \quad\mathrm{vs}   \ 0.309867\ \mathrm{for\ lattice.} \\
\end{eqnarray*}

\subsection{Lipschitz equivalence}

How does the change of the cutoff affect the geometry defined by
$\sigma$? We now show that for the free field, there are constants
$k_{1},k_{2}$ such that

\begin{equation}
0<k_{1}(a,b)\leq\frac{\sigma_{a}(x,x')}{\sigma_{b}(x,x')}\leq k_{2}(a,b)<\infty\label{eq:bi-Lipschitz}
\end{equation}

\begin{figure}
\begin{center}
\includegraphics[scale=0.6]{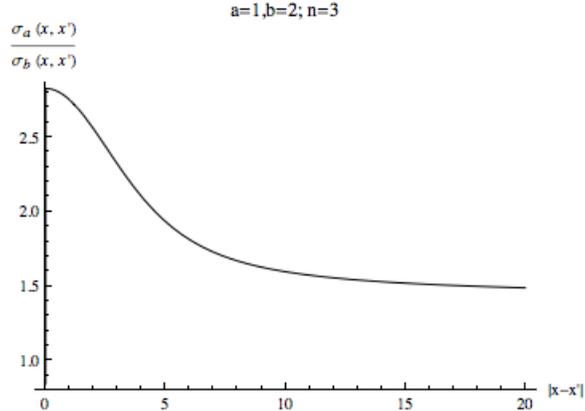}
\end{center}
\caption{\label{ratio} This plot shows the ratio of distance using two cutoffs on the Euclidean distance.}
\end{figure}

That is, $k_{1},k_{2}$ depend on the cutoffs $a,b$ but not on the
points $x,x'\in\mathbb{R}^{n}$. This means that the two metrics $\sigma_{a}$
and $\sigma_{b}$ on $\mathbb{R}^{n}$ are Lipschitz equivalent. The
proof for the free massless theory uses the explicit formula to show
that when $a<b$, the ratio $\frac{\sigma_{a}(x,x')}{\sigma_{b}(x,x')}$
takes its largest value for $x=x'$ and its smallest value as $|x-x'|\to\infty$.
Figure~\ref{ratio} illustrates this fact. Thus,

\[
\left(\frac{b}{a}\right)^{\frac{n}{2}-1}\leq\frac{\sigma_{a}(x,x')}{\sigma_{b}(x,x')}\leq\left(\frac{b}{a}\right)^{\frac{n}{2}},\quad a<b,\quad n>2
\]

We conjecture that this bi-Lipschitz inequality (\ref{eq:bi-Lipschitz})
holds for all renormalizable scalar QFT in the continuum regularization
schemes. The actual values of the Lipschitz constants might change,
however. Thus, we propose that although the metric itself depends
on the cutoff, its Lipschitz equivalence class is universal. This
makes some sense as Lipschitz equivalence for metric spaces is analogous
to diffeomorphisms for manifolds.

\section{Beyond Free Fields: Wilson-Fisher }

When $3\,<\,n\,<\,4$, the scalar field theory with potential
\begin{equation}
V(\phi)=\frac{1}{2}m^{2}|\phi|^{2}+\frac{1}{4}\lambda|\phi|^{4}
\end{equation}

has a fixed point of the renormalization group. The momentum integrals
defining the scalar theory make sense even for fractional values of
$n$, even though the case $n=3$ is the case of physical interest.
The case $n=3, N=1$ ($N$ being the number of components of $\phi$), for example, describes the critical point of a
liquid and a gas. The connection of this to fractals remains mysterious.
A more detailed study of quantum field theory on fractals is called
for. We took a first step in this direction ourselves \cite{FractalQFT}.

At this Wilson-Fisher fixed point, the Green's function is
\[
G(x-x')=\int\frac{e^{ip\cdot(x-x')}}{p^{2-\eta}}K\left(a|p|\right) \frac{d^{3}p}{(2\pi)^{3}}
\]
where the critical exponent $\eta$ can be calculated in the $\epsilon$-expansion
(Sec. 25.5 of Ref. \cite{ZinnJustin})
\begin{equation}
\eta=\frac{N+2}{2(N+8)^{2}}\epsilon^{2}+\mathrm{O}\left(\epsilon^{3}\right),\quad\epsilon=4-n.
\end{equation}

This quantity is universal and has been calculated to much higher
precision. (See \cite{Kleinert} for the result up to order $\epsilon^{5}$).
In this example, the metric
\[
\sigma^{2}(x,x')=2G(0)-2G(x-x')
\]

becomes for $|x-x'|>>a$,
\begin{equation}
\sigma(x,x')=\frac{\sigma_{3}}{a^{\frac{1+\eta}{2}}}-\frac{\sigma_{3}'}{|x-x'|^{1+\eta}}+\cdots
\end{equation}

That is, even when $n=3$, it scales as if the dimension of space
were a little bit higher than three. Again, the diameter of space
is finite and the next-to-leading order correction contains the Green's
function of physical interest. Also, for small $|x-x'|$ our $\sigma(x,x')$
is proportional to the Euclidean metric.

\section{Embedding in a Riemannian manifold}

Our standard deviation is \emph{not }a geodesic
metric. The length metric of $\sigma$ is proportional to the Euclidean
metric, which is typically larger than $\sigma$. In the lattice regularization,
the length of any curve is just the number of edges along it: the
supremum above is achieved when each segment connects nearest neighbors.
In the heat kernel regularization, we saw that when $x,x'$ are close
enough, $\sigma(x,x')$ is proportional to $|x-x'|$. So the length
of any curve according to $\sigma$ will be, up to a constant multiple,
its Euclidean length.

Thus, the distance perceived by a quantum model of propagation is
drastically different from the classical model. Classically, the particle
simply takes the shortest path (which is also the path of least action).
Quantum mechanically, we must sum over all the paths; longer paths
are simply less probable. Long paths can dominate the sum, if the
sheer number of long paths makes up for their smaller probability.
This is what happens on cubic lattices of dimension $n\geq2$ as is easy
to verify using the explicit form we found above. Again, this illustrates
how far from being a geodesic metric $\sigma$ is.

We will now show that we can embed $\mathbb{R}^n$ into a Riemannian manifold of one dimension higher such that the length of the geodesic connecting $(x,x')$ in the ambient space is equal to $\sigma(x,x')$. The extra dimension provides a ``short cut" that allows us to realize our metric as a geodesic distance. In the Appendix we give an example involving the chord length of circles that illustrates this situation.

Consider a Riemannian metric on $\mathbb{R}^{n}\times \mathbb{R}^{+}$

\begin{equation}
ds^{2}=h^{2}(\rho)d\rho^{2}+\rho^{2}dx^{i}dx^{i}\label{eq:AmbientMetric}
\end{equation}

These coordinates are chosen to make later expressions simpler.

As an example, $n=1$ and $h(\rho)=\frac{1}{\rho}$ is one description
of the metric of constant negative curvature on a hyperboloid. The
substitution $y=\frac{1}{\rho}$ turns this into the familiar Poincare
metric
\[
ds^{2}=\frac{dy^{2}+dx^{2}}{y^{2}}.
\]

\begin{figure}
\begin{center}
\includegraphics[scale=0.6]{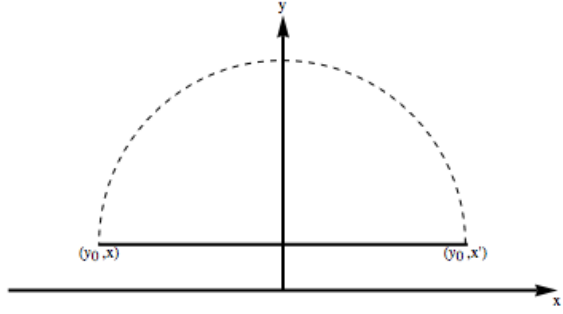}
\end{center}
\caption{\label{embed} The trajectory in the $x\,y$ plane is shown by dotted line.}
\end{figure}

To continue with this example (which we will not use directly, but
is similar enough to those we will use), the real line is a submanifold,
the line of constant $y=y_{0}$. There are two induced distances on
this submanifold: we can ask for the minimum length of curves that
lie on the submanifold- this is just the Euclidean metric on the real
line or the minimum over all curves that start and end at points
$(y_{0},x)$ and $(y_{0},x')$ on the line, but in between can lie
anywhere on the plane (Fig.~\ref{embed}). It is clear that the latter can be smaller
than the Euclidean distance. It is, in fact, a non-geodesic metric on
the real line, whose length metric is the Euclidean metric.

The geodesic (\ref{eq:AmbientMetric}) minimizes
\begin{equation}
\int\sqrt{h^{2}(\rho)+\rho^{2} \left(\frac{dx}{d\rho}\right)^{2}}d\rho
\end{equation}

We need to find the geodesic that connects the $(\rho_{0},x')$ with
$(\rho_{0},x)$. By a rotation and translation, we can choose $x=(\frac{r}{2},0,\cdots0)$
and $x'=(-\frac{r}{2},0,\cdots0)$, where $r=|x-x'|$ is the Euclidean
distance. The geodesic will then lie in the $(\rho,x^{1})$ plane.
The Euler-Lagrange equation implies that
\[
\frac{\rho^{2}}{\sqrt{h^{2}(\rho)+\rho^{2}\left(\frac{dx}{d\rho}\right)^{2}}}\left(\frac{dx}{d\rho}\right)=\rho_{1}
\]
for some constant $\rho_{1}$. A moment's thought will show that the
geodesic is reflection symmetric around $x=0$ and that $\frac{d\rho}{dx}=0$
at $x=0$. Thus  $\frac{dx}{d\rho}=\infty$ at $x=0$, and we conclude from the above
equation that $\rho_{1}$ is simply the point at which $x=0$,
\begin{equation}
\frac{dx}{d\rho}=\pm\frac{\rho_{1}h(\rho)}{\rho\sqrt{\rho^{2}-\rho_{1}^{2}}}
\end{equation}

Thus
\[
x(\rho)=\pm\int_{\rho_{1}}^{\rho}\frac{\rho_{1}h(\rho)d\rho}{\rho\sqrt{\rho^{2}-\rho_{1}^{2}}}
\]

It follows from
\[
h^{2}(\rho)+\rho^{2} \left(\frac{dx}{d\rho}\right)^{2}=\frac{h^{2}(\rho)\rho^{2}}{\rho^{2}-\rho_{1}^{2}}
\]

that
\begin{equation}
r=2\int_{\rho_{1}}^{\rho_{0}}\frac{\rho_{1}h(\rho)d\rho}{\rho\sqrt{\rho^{2}-\rho_{1}^{2}}}\label{eq:rSoln}
\end{equation}

\begin{equation}
L=2\int_{\rho_{1}}^{\rho_{0}}\frac{\rho h(\rho)d\rho}{\sqrt{\rho^{2}-\rho_{1}^{2}}}\label{eq:LSoln}
\end{equation}

Together Eqns. (\ref{eq:rSoln}) and (\ref{eq:LSoln}) give parametrically,
via $\rho_{1}$ the length of the geodesic in terms of the Euclidean
distance $r$. When $r\to0$, we have also $\rho_{0}\approx\rho\approx\rho_{1}$
and

\begin{equation}
L\approx\rho_{0}r\label{eq:LShort}
\end{equation}

The inverse problem of determining $h(\rho)$ given $L(r)$ looks
hard. But recall that only the asymptotic behavior of $\sigma(r)$
as $r\to\infty$ is universal. So we should be able to find an $h(\rho)$
within the same universality class by looking at the asymptotic behavior.
What should $h(\rho)$ be in order that
\[
L(r)\sim C-\frac{C'}{r^{n-2+\eta}}+\ldots
\]

This is the behavior of the metric $\sigma(x,x')$ in the continuum
regularization we derived in the last section.

Equation (\ref{eq:LShort}) determines $\rho_{0}$ in terms of
the cutoff.
\begin{equation}
\rho_{0}=\frac{2^{\frac{-n}{2}}\pi^{-\frac{n}{4}}}{\sqrt{n}a^{\frac{n}{2}}}
\end{equation}

Suppose $h(\rho)\sim h_{1}\rho^{-\mu}$ as $\rho\to0$. Then we get
\begin{eqnarray*}
r&\approx&2h_{1}\rho_{1}\frac{\rho_{1}^{-1-\mu}-\rho_{0}^{-1-\mu}}{\mu+1} \\
r&\approx&\frac{2h_{1}}{\mu+1}\rho_{1}^{-\mu}\\
\end{eqnarray*}

Also,
\[
L\approx2h_{1}\frac{\rho_{0}^{1-\mu}-\rho_{1}^{1-\mu}}{1-\mu}
\]

so that
\[
L\approx\frac{2h_{1}}{1-\mu}\rho_{0}^{1-\mu}-\frac{2h_{1}}{1-\mu}\left[\frac{\mu+1}{2h_{1}}\right]^{\frac{\mu-1}{\mu}}r^{1-\frac{1}{\mu}}
\]

This gives us what we want if
\begin{eqnarray*}
1-\frac{1}{\mu}&=&-(n-2+\eta)\\
\mu&=&\frac{1}{n-1+\eta}\\
\end{eqnarray*}

The resulting metric
\[
ds^{2}\approx h_{1}^{2}\rho^{-\frac{2}{n-1+\eta}}d\rho^{2}+\rho^{2}dx^{i}dx^{i}
\]
has curvature going to $-\infty$ as $\rho\to0$, when $(n+\eta-1)\geq1$.
The case $n=2,\eta=0$ is marginal. In that we get a metric of constant
negative curvature asymptotically.

\section{Conclusions and Further Directions}

Our main point is that a non-Riemannian metric on space-time emerges from scalar  quantum field theory. In dimensions $n>2$, even the free field induces a very different metric from Euclidean space: space-time has finite diameter, for example. Yet, the length of any curve as  defined by this metric is (up to a constant) the usual Euclidean length. Thus,  classical measurements are unaffected. We calculated the metric explicitly in free field theory and also took a step toward understanding the interactions by calculating it for the Wilson-Fischer fixed point. It would be of interest to also study the case of $\lambda\phi^6$ interactions, as they  describe multicritical points and marginal perturbations.

It is of great interest to calculate the metric (perhaps exactly) for the case of asymptotically free scalar quantum field theories in two dimensions (e.g., the nonlinear sigma model). We would expect that there are logarithmic corrections to the length of a curve, a first  indication of non-Riemannian geometry. Also, there are many two-dimensional scalar  field theories that are exactly solvable; can we get an exact formula for the metric in some of them?

But by far the question of greatest interest is that of $\lambda\phi^4$ theory in four dimensions. Since perturbation theory breaks down at short distances, it is not possible to study this question analytically. A numerical simulation is needed to understand how $\sigma(x,x')$ depends on, or differs from,  the Euclidean distance $|x-x'|$. Do they even define the same topology? Is the Euclidean length still the length induced by $\sigma$? How does the embedding in section 7 of two-dimensional space-time into three-dimensional hyperbolic space change in the presence of interactions? Is there a connection to the AdS/CFT correspondence?

These questions are especially urgent in view of the expected discovery of the Higgs boson at the LHC. If such a fundamental scalar field exists, and there is no evidence of supersymmetry, the metric geometry induced by it might play a role in understanding the hierarchy problem of the standard model. In any case, as a natural property of the scalar quantum field,  it is  of interest to study $\sigma$  in  numerical simulation of lattice scalar field theory.

\section{Acknowledgements}

We especially thank L. Gross and R. Strichartz for explanations of metric geometry and $L^p$ averages.
We thank  also  A. Iosevich, A. Joseph, Y. Meurice,  F. Moolekamp, E. Prassidis,
and B. Ugurcan for discussions. A. K. was supported
in part by a grant from the U.S. Department of Energy under Contract No.
DE-FG02-91ER40685.

\section{Appendix : Metric Geometry}

A metric on a set $X$ is a function $d:X\times X\to\mathbb{R}$
such that
\begin{itemize}
\item $d(x,x)=0$
\item $d(x,x')>0$ if $x\neq x'$ , separation
\item $d(x,x')=d(x',x)$, symmetry
\item $d(x,x')\leq d(x,x'')+d(x'',x')$, the triangle inequality.
\end{itemize}

The most familiar example is the Euclidean metric on $\mathbb{R}^{n}$.
\[
|x-x'|=\sqrt{\sum_{i=1}^{n}(x^{i}-x'^{i})^{2}}
\]

This metric is so ingrained in us that we might forget that the actual
metric of space-time should be deduced by physical measurements and
is not self-evidently Euclidean. Often (e.g., numerical simulations
of scalar field theory, solution of partial differential equations), we have to approximate space
by a discrete lattice $\Omega_{a,L}^{n}=a\left(\mathbb{Z}/\Lambda\mathbb{Z}\right)^{n}$
with nearest-neighbor spacing $a$ and period $L=\Lambda a$ in each
direction. Then the Euclidean metric is approximated by the length
of the shortest path connecting two points on the lattice

\begin{equation}
l(x,x')=\sqrt{\sum_{i=1}^{n}(x^{i}-x'^{i}\ \mathrm{mod}\ L)^{2}}.
\end{equation}

The square root of a metric is again a metric. More generally, if $f:\mathbb{R}^+\to\mathbb{R}^+$ is a concave function $f''(x)<0$ with $f(0)=0$, then we can construct from a given metric $d$, a new one $\tilde{d}(x,x')=f((d(x,x'))$. But the square of a metric is not always a metric. For example, the square of the Euclidean metric is not a metric: the sum of the squares of the sides of an obtuse triangle is not greater than the square of the side opposite.

The length of the shortest curve (geodesic) connecting
two points in a Riemannian manifold is a metric. Any metric arising
as the length of geodesics has the intermediate property: given any pair
of points $(x,x')$ there is another $x''$ (a midpoint) that saturates
the triangle inequality:
\[
d(x,x')=d(x,x'')+d(x'',x')
\]

Any point $x''$ lying along the shortest geodesic connecting $x,x'$
would suffice. There are metrics that do not have this property. According
to them, the distance between two points can be shorter than the length
of every curve connecting them. Obviously, such metrics are non-Riemannian.
This is precisely the case of interest to us.

Although the concept of derivative
does not make sense, in general, on a metric space (for that we would
need a differential manifold), Lipschitz functions are the analogue
of differentiable functions. A function $f:X\to Y$ between metric
spaces is said to be $k$-Lipschitz if
\[
\frac{d_{Y}(f(x),f(x'))}{d_{X}(x,x')}<k
\]
for all $x,x'\in X$. Roughly speaking, the magnitude of the derivative
is less than $k$. Two metric spaces are Lipschitz equivalent if there
are continuous, one-to-one Lipschitz maps in each direction which are
inverses of each other. Lipschitz equivalence is roughly analogous
to diffeomorphisms between manifolds.

\subsection{Length of curves}

Given a metric we can define the length of a curve as the largest sum
of the length of line segments. In more detail, a curve $\gamma:[0,T]\to X$
can be broken up into segments
\[
0\equiv t_{0}<t_{1}<t_{2}<\cdots t_{k}<T\equiv t_{k+1}
\]

The sum of the chord lengths
\[
\sum_{i=1}^{k+1}d(\gamma(t_{i-1}),\gamma(t_{i}))
\]

can be thought of as an approximation to its length. The actual length
of the curve is the least upper bound of all such approximations;
i.e., avoiding all the ``short cuts'' made by the chords:
\[
l[\gamma]=\sup_{0<t_{1}<\cdots t_{k}<T}\sum_{i=1}^{k+1}d(\gamma(t_{i-1}),\gamma(t_{i}))
\]

The length of a continuous curve can be infinite. There are well-known examples of continuous curves (e.g., Koch curve) with an infinite length in the Euclidean metric. Also, suppose we define $d(x,x')=|x-x'|^{1\over 2}$, the square root of the Euclidean distance. Then the length of every  straight-line segment  is infinite!

For the familiar case of a differentiable curve in a Riemannian manifold,
it is not hard to verify that this agrees with the usual definition
\begin{equation}
l[\gamma]=\int_{0}^{T}\sqrt{g_{\gamma_{t}}\left(\dot{\gamma(t)},\dot{\gamma(t)}\right)}dt
\end{equation}

The reason is that, for short-enough segments, the chord length is
approximated by the length of the tangent vector. It is not hard to
come up with continuous but not differentiable curves of infinite
length.

\subsection{Length metric}

Given a metric $d$, we can often construct from it a (possibly distinct)
length metric $d_{l}(x,x')$ as the greater lower bound of the lengths
of all the curves that connect $x$ to $x'$. (This construction could fail if the length of every continuous curve is infinite, or if there is no greater lower bound.)

A metric is said to
be \emph{geodesic} (also called an interior space or intrinsic metric)
if this is the one we started with: $d_{l}(x,x')=d(x,x')$.

The Euclidean distance is an example of a geodesic metric. Any length
metric is, itself, a geodesic metric. That is, $(d_{l})_{l}=d_{l}$
for any $d$. For more on these matters see Ref. \cite{Burago}.

\begin{figure}
\begin{center}
\includegraphics[scale=0.5]{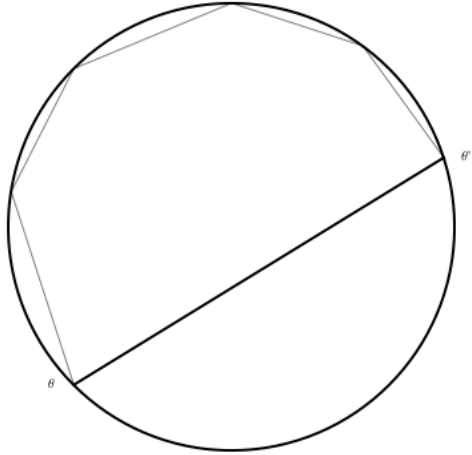}
\end{center}
\caption{\label{chord} The diagram shows how tangents are approximated by chord lengths.}
\end{figure}

An example of a nongeodesic metric (Fig.~\ref{chord}) is the  length of a chord connecting two points on a circle:
\[
d(\theta,\theta')=2\sin{|\theta-\theta'|\over 2}
\]

where $0\leq \theta \leq2\pi $ is the usual polar coordinate. For small angles this agrees with the arc length
\[
d(\theta,\theta')\approx |\theta-\theta'|.
\]

Therefore, if we break up an arc into small segments and add up the lengths, we will get the arc length; i.e., the length metric of $d$ is just the arc length
\[
d_l(\theta,\theta')=|\theta-\theta'|.
\]

But, in general,
\[
d(\theta,\theta')<d_l(\theta,\theta')
\]

Note that although the chord length is not a geodesic metric on the circle, it is the length of a geodesic in  the plane in  which the circle is embedded. We showed that the standard deviation metric of a scalar field theory can be similarly realized as the geodesic length in a space of one dimension higher.

\subsection{Triangle inequality for standard deviation metric}

Suppose that $a_{i},b_{i},c_{i}$ (for some finite range of the index
$i$) are positive numbers satisfying the inequality $a_{i}\leq b_{i}+c_{i}.$
Then it is obvious that the weighted averages $\langle a \rangle=\frac{\sum_{i}a_{i}w_{i}}{\sum_{i}w_{i}}$
also satisfy $\langle a \rangle \leq\langle b \rangle+\langle c \rangle.$ More generally, the $L^{p}$ averages
for $p\geq1,$
\begin{equation}
\langle a \rangle_{p}=\left[\frac{\sum_{i}a_{i}^{p}w_{i}}{\sum_{i}w_{i}}\right]^{\frac{1}{p}}
\end{equation}
satisfy
\[
\langle a \rangle_{p}\leq\langle b \rangle_{p}+\langle c \rangle_{p}.
\]

To see this, simply note that
$\langle a \rangle_{p}\leq \langle b+c\rangle_{p}$ by monotonicity; the rest follows by the fact
that the $L^{p}$ norm satisfies the triangle inequality.

If we replace the discrete average above by an integral with respect to a probability measure $e^{-S(\phi)}d\phi$, the inequality continues to hold. For positive functions,
\[
a({\phi })\leq b({\phi})+c({\phi })\implies \langle a \rangle_{p}\leq\langle b \rangle_{p}+\langle c \rangle_{p}.
\]
\begin{equation}
\langle a \rangle_{p}=\left[\frac{\int e^{-S(\phi)}a^p({\phi })d\phi }{\int e^{-S(\phi)}d\phi }\right]^{\frac{1}{p}}
\end{equation}

These facts are useful for us because they show that the $L^p$ average of a metric is also a metric. We just have to choose
\[
a(\phi)=|\phi(x)-\phi(x')|, b(\phi)=|\phi(x)-\phi(x'')|,c(\phi)=|\phi(x'')-\phi(x')|
\]
Thus $\sigma(x,x')=\sqrt{\langle(\phi(x)-\phi(x'))^2\rangle}$ satisfies the triangle inequality.

We thank L. Gross for illuminating this point.

\end{document}